\documentclass[letterpaper]{jpconf}
\usepackage{graphicx}
\input babarsym

\begin{document}
\title{Bottomonium at \babar}

\author{Francesco Renga (for the \babar\ Collaboration)}

\address{``Sapienza'' Universit\`a di Roma and INFN Roma}

\begin{abstract}
Originally designed for CP violation studies in the \B meson system, the \B-Factories recently showed an exciting capability for improving
our experimental knowledge in the field of hadron spectroscopy. Here I will present some of the most recent \babar\ results
concerning bottomonium spectroscopy. In particular, I'll report the first observation of the ground state $\eta_b$ in $\Upsilon(nS) \to
\gamma \eta_b$ and the results of an energy scan in the range of 10.54 to 11.20\gev, that produced a new measurement of
the $\epem \to \bbbar$ cross section in the region of the \FourS and candidate \FiveS and \SixS resonances, with an integrated
luminosity 30 times larger than the previous scans.
\end{abstract}

\section{Introduction}

In the last few years, quarkonium spectroscopy received significant
contributions from the \B-Factory experiments \babar\ and Belle.
This impact have been recently boosted by the decision of the \babar\ 
Collaboration~\footnote{A description of the \babar\ detector can be found elsewhere~\cite{NIM}.}
of running  the PEP-II \B-Factory at different Center of Mass (CM) energies, with the main goal 
of investigating bottomonium properties at a deeper level. About $\sim 28 \invfb$ have been collected at the \ThreeS resonance, 
providing the largest sample available worldwide at this CM energy.
A sample of $\sim 14.5 \invfb$ has been collected at the \TwoS resonance, 
and an energy scan of the region above the \FourS resonance has been performed.

Here I will report some of the first results obtained in these unique samples:
the discovery of the $\eta_b$ in $\ThreeS \to \gamma \eta_b$~\cite{etab}, confirmed in 
$\TwoS \to \gamma \eta_b$~\cite{etab_2S}, and a measurement of the inclusive cross section $\sigma(\epem \to \bbbar)$ in the range of 10.54 to 11.20\gev~\cite{scan}. 

\section{The $\eta_b$ Discovery}

The $\eta_b(1S)$ (simply $\eta_b$ hereafter) is the ground state of the bottomonium spectrum, discovered by the
\babar\ collaboration in the $\ThreeS \to \eta_b \gamma$ decay channel, by exploiting a sample of $(109 \pm 1)$
million of \ThreeS. The mass of the $\eta_b$ was expected to lie around 9.4\gevcc, hence the
analysis consists of the search for a monochromatic photon of about 900\mev in the \ThreeS rest frame, accompanied by a set of
charged tracks and electromagnetic clusters consistent with a hadronic $\eta_b$ decay.

Photons are identified as calorimeter clusters isolated from tracks and with a shape consistent with an
electromagnetic shower, by requiring a lateral momentum~\cite{LAT} less than 0.55.
A \piz veto is also applied, by rejecting photons that, combined with other neutral clusters in the event, give an invariant mass
consistent with a \piz hypothesis within 15\mevcc. In order to achieve a better resolution and a lower background, only the central part of
the electromagnetic calorimeter ($0.762 < \cos(\theta_{\gamma,\,LAB}) < 0.890$) is used in this analysis. Hadronic $\eta_b$ decays are
selected by requiring at least four tracks in the event. In order to reject the QED background, we require
the ratio $R_2$ between the $0^{th}$ and $2^{nd}$ order Fox-Wolfram moments~\cite{FW} to be less than 0.98. A selection is finally
applied on the angle between the photon and the $\eta_b$ thrust axis~\cite{thrust,thrust_2}. After this selection, the background is
composed of a non-peaking contribution from light mesons decays
and peaking contributions from the initial state radiation (ISR) process $\epem \to \gamma_{ISR}\OneS$
and the bottomonium transitions $\chi_{bJ}(2P) \to \gamma \OneS$ ($J = 0,\,1,\,2$).

\begin{figure}[h]
\begin{center}
\includegraphics[width=14pc]{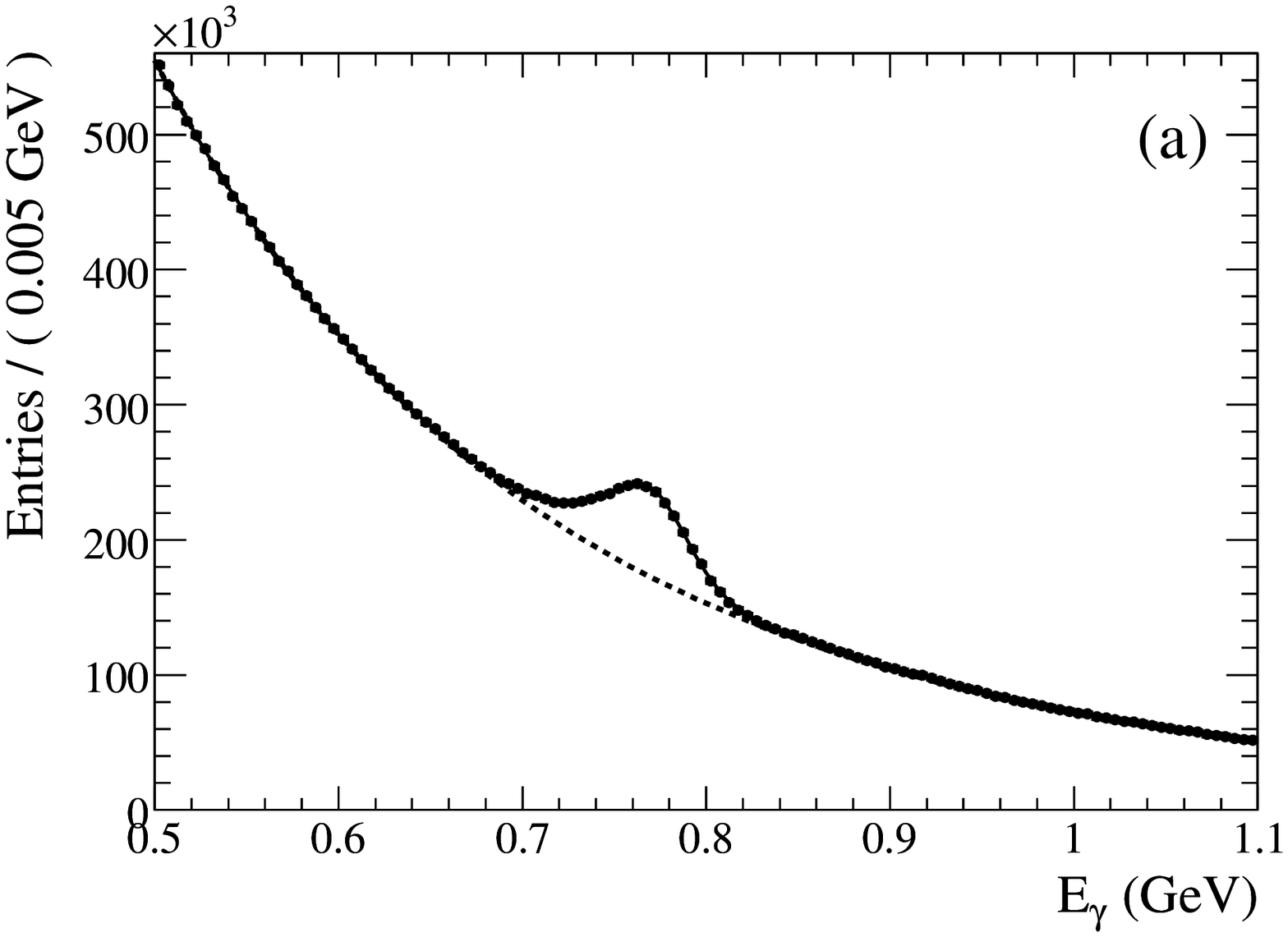}\hspace{2pc}
\includegraphics[width=14pc]{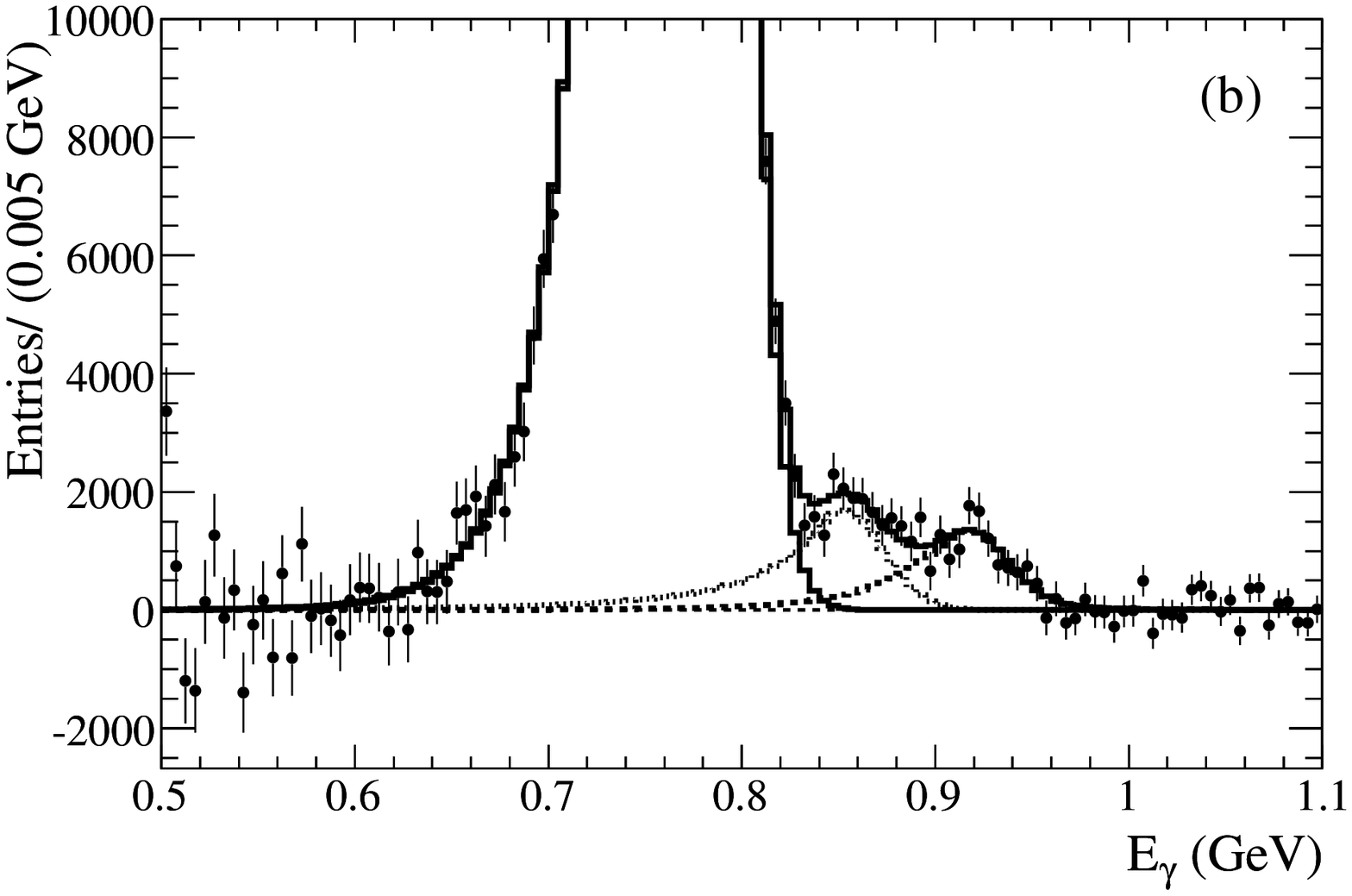}
\caption{\label{fig:egamma} (a) Spectrum of $E_\gamma$. The dashed line show the non-peaking background component.
(b) Spectrum of $E_\gamma$ after subtracting the non-peaking background component, with PDFs for 
$\chi_{bJ}(2P)$ peak (solid), ISR \OneS (dot), $\eta_b$ signal (dash) and 
the sum of all three (solid).}
\end{center}
\end{figure}

In figure~\ref{fig:egamma} the photon spectrum after the selection is shown. A binned maximum likelihood (ML) fit of the
spectrum is performed in the region between 0.5 and 1.1\gev with four components: non-peaking background, 
$\chi_{bJ}(2P) \to \gamma \OneS$, $\gamma_{ISR}\OneS$ and the $\eta_b$ signal. The non-peaking background is parameterized with a
probability density function (PDF) given by $\mathcal{P}(E_\gamma) = A\left(C + \exp\left[ -\alpha E_\gamma - \beta E_\gamma^2 \right]
\right)$. The $\chi_{bJ}(2P) \to \gamma \OneS$ background is described by the superposition of three Crystal Ball (CB) PDFs~\cite{CB}, one
for each $J$ state. The ISR background is parameterized by a single CB PDF while the signal is described by the convolution of a
Breit-Wigner and a CB PDF. The photon spectrum after non-peaking background rejection is also shown in figure~\ref{fig:egamma}. 
The fit yields $19200 \pm 2000 \pm 2100$ signal events, corresponding to $\BR(\ThreeS \to \eta_b
\gamma) = (4.8 \pm 0.5 \pm 1.2) \times 10^{-4}$. A significance of more than 10 standard deviations has been associated to this 
signal.

The measured $\eta_b$ mass is $(9388.9^{+3.1}_{-2.3} \pm 2.7)\mevcc$, corresponding to a hyperfine splitting of
$M(\OneS) - M(\eta_b) = (71.4_{+3.1}^{-2.3} \pm 2.7)\mevcc$. It is in agreement with recent lattice results~\cite{lattice}, but a
significant disagreement is found with respect to QCD calculations~\cite{NRQCD}.

\begin{figure}[h]
\begin{center}
\includegraphics[width=14pc]{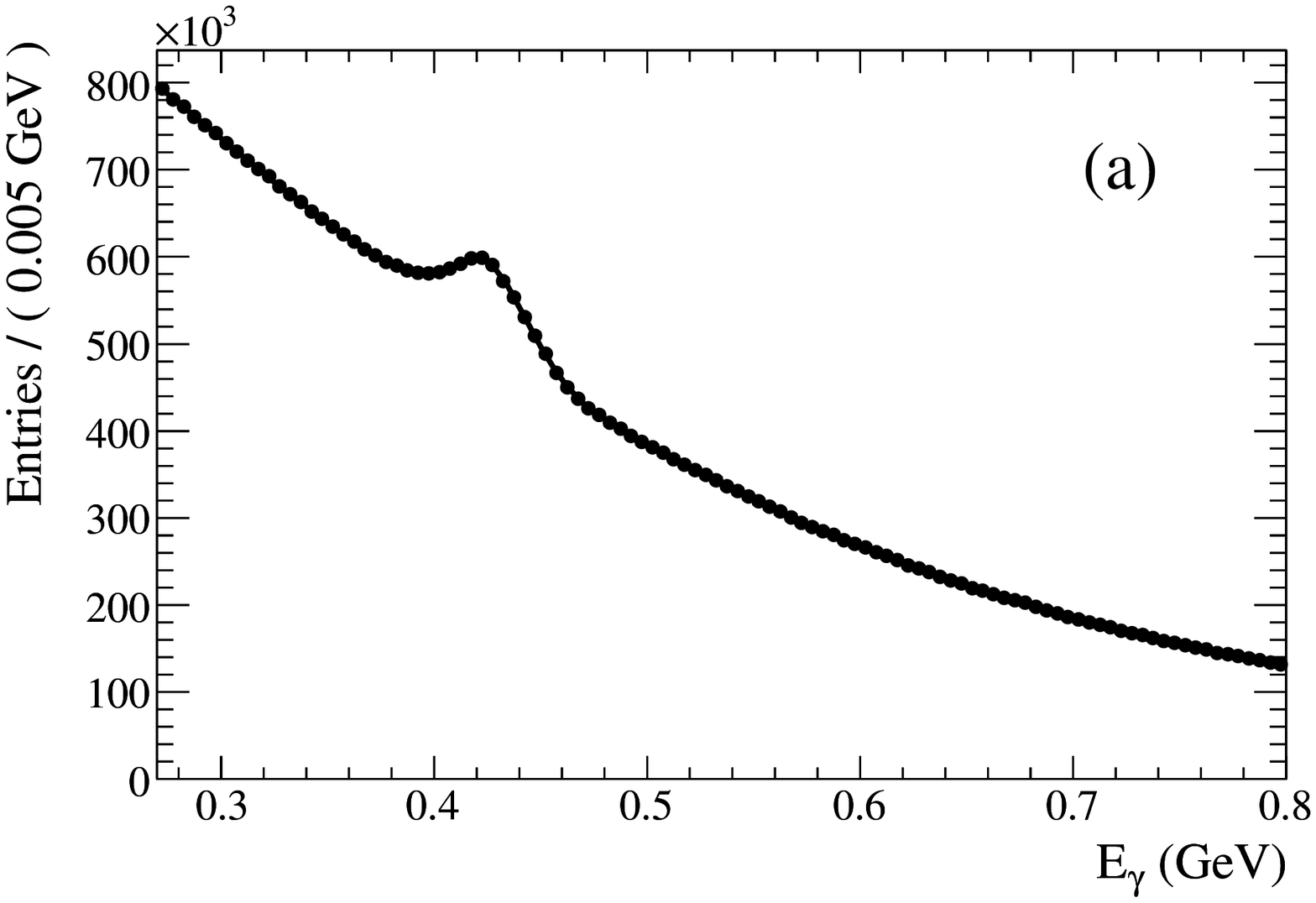}\hspace{2pc}
\includegraphics[width=14pc]{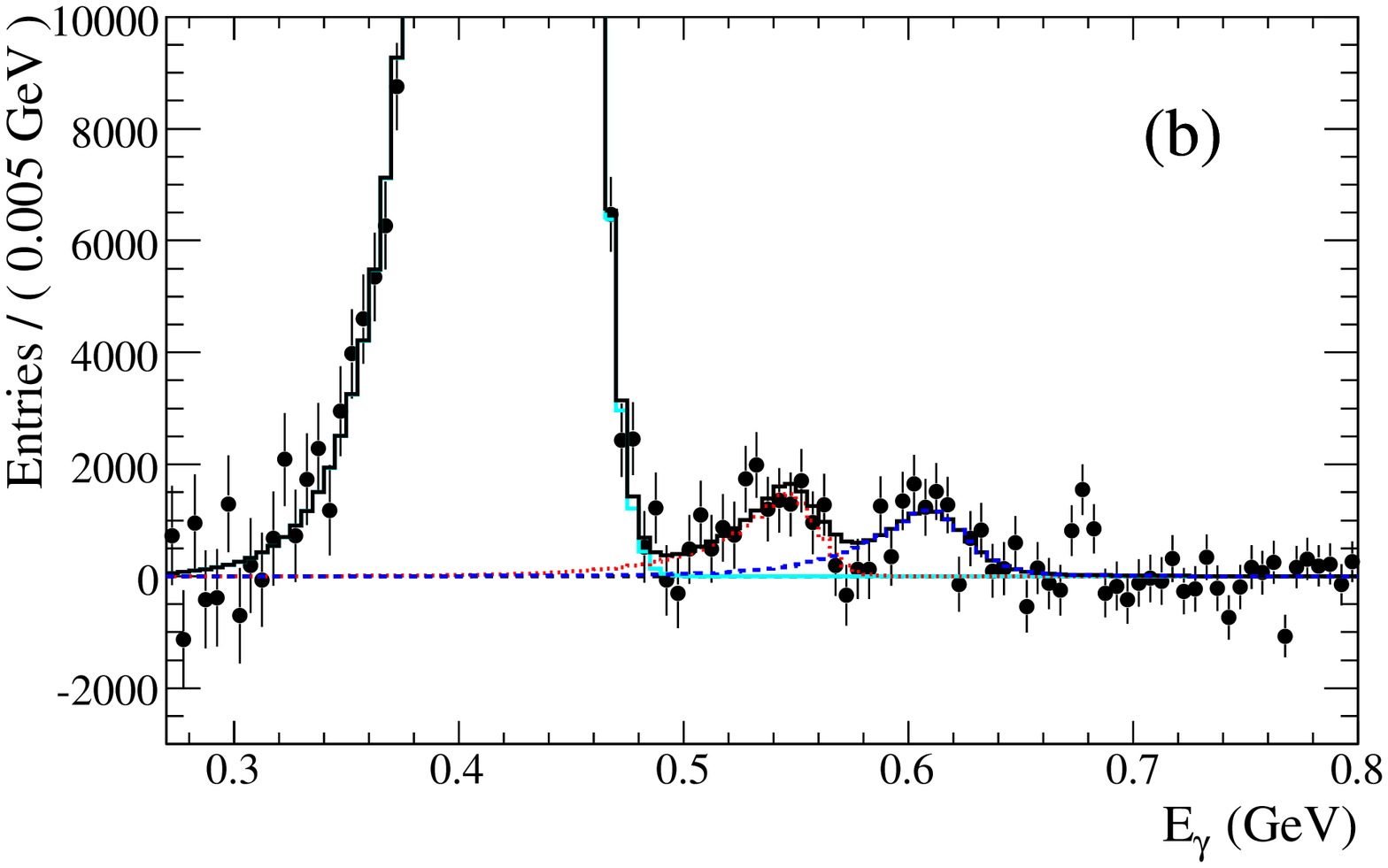}
\caption{\label{fig:egamma_2S} (a) Spectrum of $E_\gamma$ in the $\TwoS \to \eta_b
\gamma$ analysis. (b) Spectrum of $E_\gamma$ after subtracting the non-peaking background component, with PDFs for $\chi_{bJ}(2P)$
peak (solid), ISR \OneS (dot), $\eta_b$ signal (dash) and the sum of all three (solid).}
\end{center}
\end{figure}

This result has been confirmed by a similar analysis performed on the \TwoS data sample, looking for $\TwoS \to \eta_b
\gamma$. In this case,  a lower energy photon is present, implying a larger non-peaking background but also a better 
absolute energy resolution, allowing for a better separation of the signal from the other peaking components. We obtained
$M(\eta_b) = (9392.9^{+4.6}_{-4.8} \pm 1.8) \mevcc$ and $\BR(\TwoS \to \eta_b \gamma) = (4.2^{+1.1}_{-1.0} \pm 0.9) \times 10^{-4}$, with
a 3.5$\sigma$ signal significance. The corresponding fit is shown in figure~\ref{fig:egamma_2S}.

\section{$\sigma(\epem \to \bbbar)$ scan above the \FourS resonance}

The recent discovery of exotic charmonium-like states~\cite{faccini} suggest the possibility of the existence of similar
bottomonium-like states. A naive scaling of the new states, according to the typical mass difference between bottomonia and charmonia,
suggests that new bottomonium states could lie in the region between the \FourS and the candidate \FiveS and \SixS.  The \babar\
Collaboration performed an energy scan of this region in order to investigate this possibility.

The CM energy $\sqrt{s}$ has been moved from 10.54 to 11.20\gev, in steps of 5\mev, collecting about 25\invpb per step, for a total of
3.3\invfb. An additional scan of the \SixS region, with 8 steps of 600\invpb has been also performed. The \babar\ scan improves by a factor of 30 the statistics of the previous scans~\cite{CLEO_scan,CUSB}, with 4 times finer steps.

We adopted an inclusive analysis strategy, looking for unexpected structures in the hadronic ratio $R_b = \sigma(\epem \to
\bbbar)/\sigma_0(\epem \to \mumu)$, where $\sigma_0(\epem \to \mumu) = 4\pi\alpha^2/3s$ is the point like $\epem \to \mumu$
cross section. We normalized our measurement to the measured $\epem \to \mumu$ cross section, by writing $R_b = k \times
N_{\bbbar}/N_{\mumu}$, where $N_{\bbbar}$ ($N_{\mumu}$) is the number of produced \bbbar (\mumu) pairs and $k$ account for radiative
corrections to the point-like \mumu cross section (estimated from MC calculation using KK2f~\cite{KK2f}). 

In order to estimate $N_{\bbbar}$ and $N_{\mumu}$, we need to reconstruct and select a \bbbar and a \mumu sample, and correct the
number of observed events for background contamination and signal efficiency. The \bbbar sample is selected by requiring at least three
tracks in the event and a reconstructed energy of at least 4.5\gev. The vertex of the tracks is required to be within 5~mm from the beam
crossing in the transverse plane and 6~cm along the beam axis. The selection requirement $R_2 < 0.2$ is also applied to reject 
the $\epem \to \qqbar$, $q=(u,d,s,c)$ background. Dimuon events are selected by requiring exactly two tracks with invariant mass 
larger than 7.5\gevcc, a polar angle $\theta < 0.7485$ and a collinearity better than 10\degrees. 

The first scan point, at 10.54\gev, where no \bbbar production is expected, is used as a reference point to evaluate the
background contaminating the \bbbar sample. Two components are present: residual \qqbar background and two-photons 
$\epem \to \gamma^\ast \gamma^\ast \epem \to X \epem$. Their cross sections are estimated at the reference point and scaled
according to the expected trend ($\sqrt{s}$ for the \qqbar background and $\log(s)$ for the two-photon background), while background
efficiencies are evaluated by means of MC simulation for different CM energies. Similar simulations are used in order to estimate the
signal efficiency.

The \mumu sample is also used for a precise measurement of the center of mass energy for each scan point, extracted by means of a fit
of the \mumu invariant mass spectrum. This strategy has been validated by using data 
collected around the \ThreeS peak and comparing the results with the most precise determination of the \ThreeS peak
position~\cite{novosibirsk}.

\begin{figure}[h]
\begin{center}
\begin{minipage}{14pc}
\includegraphics[width=14pc]{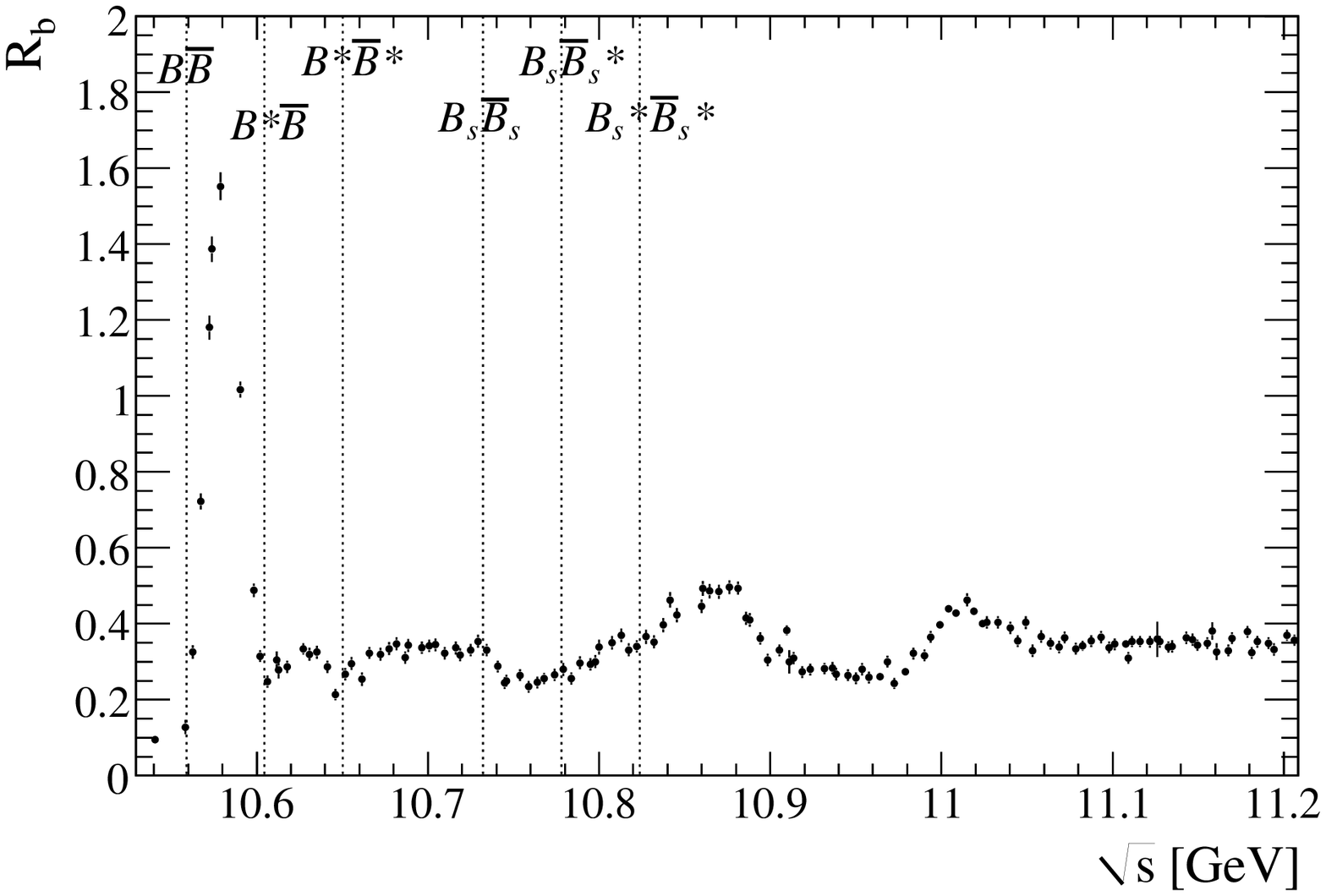}
\caption{\label{fig:scan}Measured $R_b$ as a function of the center of mass energy $\sqrt{s}$, with the position of
  the $\epem \to B^{(\ast)}_{(s)} \overline B^{(\ast)}_{(s)}$ thresholds.}
\end{minipage}\hspace{2pc}%
\begin{minipage}{14pc}
\vspace{-2.5pc}\includegraphics[width=14pc]{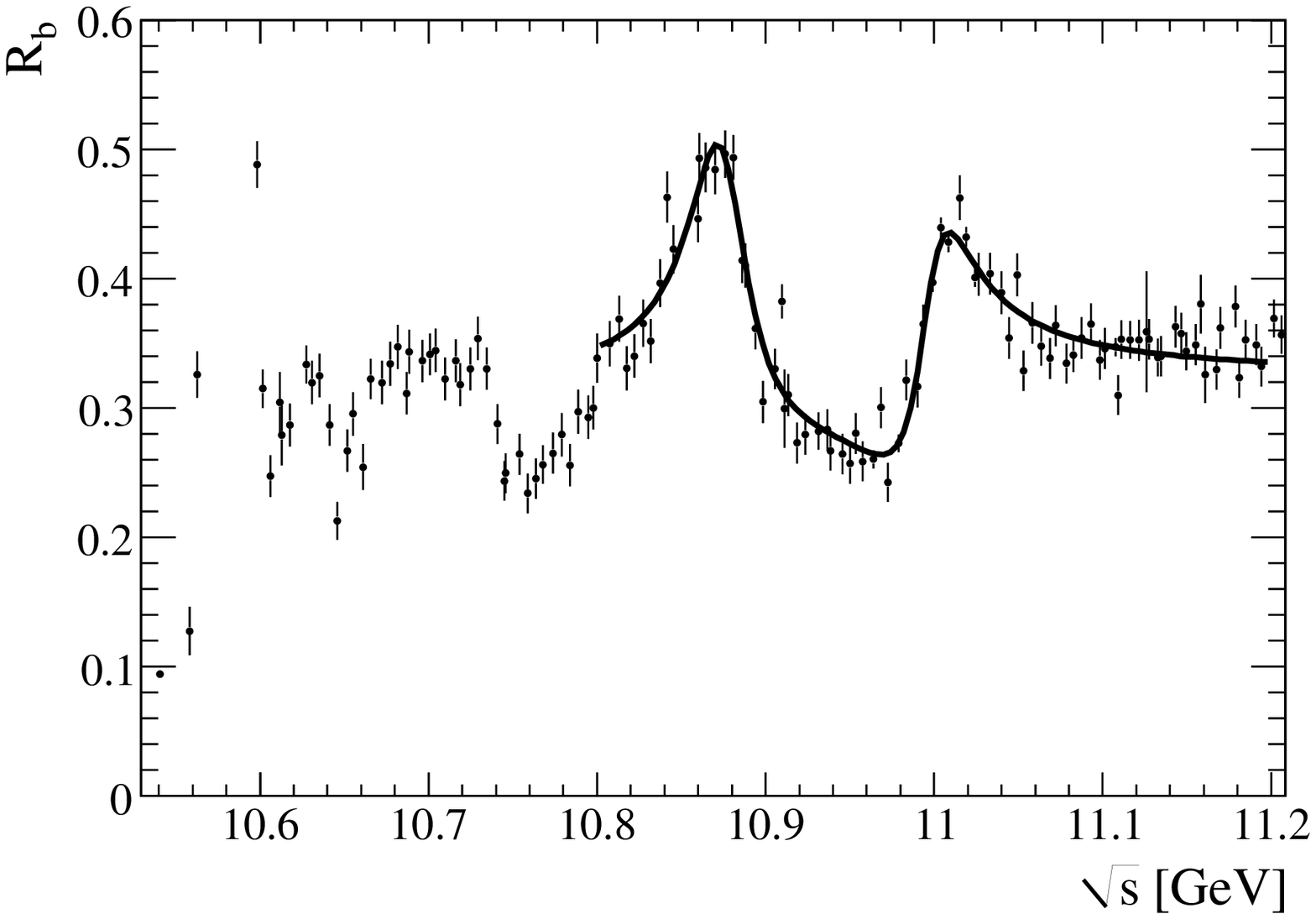}
\caption{\label{fig:fit}Fit of the \FiveS and \SixS resonance shapes.}
\end{minipage} 
\end{center}
\end{figure}

Figure~\ref{fig:scan} shows the result of the scan. The presence of several thresholds in the explored region makes difficult the
interpretation of the results. Two evident structures between 10.60 and 10.75\gev are present, in agreement with theoretical
predictions~\cite{tornqvist}. A fit for the extraction of the \FiveS and \SixS masses and widths has been also performed and 
is shown in figure~\ref{fig:fit}. The two resonances are modeled with two Breit-Wigner functions and a flat continuum is added. These
components are also allowed to partially interfere. The results are quoted in Table~\ref{tab:fit} and show a significant disagreement 
with respect to the present world averages. Anyway, we want to stress that such a kind of naive parameterization is not suitable for
describing the \bbbar production near threshold, and more refined models should be used~\cite{eichten}. The disagreement between the
current world average and our result, which is based on a more detailed scan, actually demonstrates that naive fits can provide inconsistent 
results and should be interpreted with care.

\begin{table}[h]
\caption{\label{tab:fit}Results of the \babar\ fit and comparison with the PDG world averages~\cite{PDG}.} 
\begin{center}
\lineup
\begin{tabular}{lllll}
\bhline
& \multicolumn{2}{c}{\babar} & \multicolumn{2}{c}{PDG} \\
&\FiveS&\SixS&\FiveS&\SixS\\
\hline
 mass (\gevcc) & $\010.876\pm0.002$& $10.996\pm0.002$& $\010.865\pm 0.008 $&$11.019\pm 0.008$\\
width (\mevcc) & $\043\pm4$&  $37\pm3$&$110\pm13$&$79\pm 16$\\
\bhline
\end{tabular}
\end{center}
\end{table}

\section*{References}                                                                                                   
\bibliography{bottomonium}                                                                                                                       

\end{document}